# All-optical electric field sensing with nanodiamond-doped polymer thin films

Roy Styles[1*], Mengke Han[2], Toon Goris[3], James Partridge[1], Brett C. Johnson[1], Blanca del Rosal[1], Amanda N. Abraham[1], Heike Ebendorff-Heidepriem[2], Brant C. Gibson[1], Nikolai Dontschuk[3], Jean-Philippe Tetienne[1*], Philipp Reineck[1*]

[1] School of Science, RMIT University, VIC, 3001
[2] University of Adelaide, Adelaide, SA 5005
[3] School of Physics, University of Melbourne, Parkville, Victoria 3010, Australia

* roy.styles@rmit.edu.au, jean-philippe.tetienne@rmit.edu.au, philipp.reineck@rmit.edu.au

**Abstract.** The nitrogen-vacancy (NV) center is a photoluminescent defect in diamond that exists in different charge states, NV$^-$ and NV$^0$, that are sensitive to the NV's nanoscale environment. Here, we show that photoluminescence (PL) from NV centers in fluorescent nanodiamonds (FNDs) can be employed for all-optical voltage sensing based on electric field-induced NV charge state modulation. More than 95% of FNDs integrated into a capacitor device show a transient increase in NV$^-$ PL intensity up to 31% within 0.1 ms after application of an external voltage, accompanied by a simultaneous decrease in NV$^0$ PL. The NV$^-$ PL signal increases with increasing electric field from 0 to 625 kV cm$^{-1}$. The electric field sensitivity of a single FND is 19 V cm$^{-1}$ Hz$^{-½}$. The NV charge state photodynamics are investigated on the millisecond timescale, and we find that the change in NV PL strongly depends on the rate of photoexcitation. We propose a model that qualitatively explains our results based on an electric field-induced redistribution of photoexcited electrons from substitutional nitrogen to NV centers, leading to a transient conversion of NV$^0$ to NV$^-$ centers. Our results contribute to developing FNDs as reliable, all-optical, nanoscale electric field sensors in solid-state systems.

**Introduction.** Nitrogen vacancy (NV) centers in diamond are emerging as powerful nanoscale sensors for magnetic fields,[1] temperature[2], and electric fields[3]. Most sensing approaches are based on the photoluminescence (PL) read-out of the NV electron spin. NV spins are highly sensitive to external stimuli, but their read-out generally requires microwave (MW) field for precise spin control that enables techniques like optically detected magnetic resonance (ODMR). The technological implementation of the MW spin control can be challenging for applications where the MW fields are strongly attenuated by the environment of the sensing target (e.g. in water), the sensing target may be sensitive to MW fields (e.g. in biology), or size weight and power (SWaP) considerations are paramount (e.g. miniaturized and integrated sensors).

An alternative approach is to use the NV charge state as an all-optical sensor. The PL properties of the different NV charge states differ significantly; while the negative (NV$^-$) and neutral (NV$^0$) charge states mostly emit in the red (650-750 nm) and orange (575-700 nm) spectral regions, respectively, the positive charge state is non-fluorescent. Hence, external stimuli that affect the NV charge state can be detected via changes in the NV PL intensity at specific wavelengths or ratiometrically at multiple wavelengths.

The NV charge state is generally not affected by magnetic fields or temperature but is sensitive to electric fields and voltages [4–10]. The electric field sensitivity is based on near-surface band bending related to the negative electron affinity of some diamond surface terminations, most notably hydrogen termination[11,12]. The band bending is associated with charge transfer between NV$^-$ centers and the hydrogen-terminated diamond surface, creating more NV$^0$ defects near the particle surface. External electric fields modulate the near-surface band bending and, hence, the NV charge state. Recently, it has been demonstrated that electric field changes affect spin measurements of NVs in in nanodiamonds[13] and of near-surface NVs in bulk diamond[14]. The connection between FND surface charges and NV spin properties may open new opportunities to probe the FND charge environment via optical spin measurements[15]

The simplicity of charge state sensing via changes in the NV spectral properties may offer advantages in electric field and voltage sensing. It has been investigated in bulk diamond [4–9] and, to a far lesser degree, in nanodiamonds[10]. Recently, substantial progress has been made in developing a voltage imaging platform based on the charge state of near-surface NV centers in bulk diamond.[5] In bulk diamond, the creation of NVs within a few nanometers of the surface is essential since the electric field-induced band bending that modulates the NV charge state is only effective within ~10 nm of the diamond surface. In FNDs, on the other hand, the fraction of NVs that are 'near-surface' can be easily tuned by the particle size.

However, to date, only a few studies have focused on charge state-based sensing using nanodiamonds [10,16,17], including pH sensing[16] and detecting the attachment of charged molecules[17]. The voltage-induced modulation of the NV charge state in nanodiamonds in a water-based electrochemical cell was reported by Karavelli et al [10]. Nanodiamonds were immobilized at the interface between a transparent conductive substrate and an electrolyte solution, and the voltage was applied with an electrode immersed in the electrolyte. The NV PL intensity of both hydrogenated and hydroxylated FNDs was modulated by an external voltage, and the observed changes in PL were attributed to voltage-induced changes in the NV charge state. Charge transfer between substrate and FNDs was identified as the cause for the observed PL signals for the hydrogenated NDs and voltage-induced band bending for the hydroxylated FNDs. Only 21% of hydroxylated FNDs showed a significant change in PL.

Here, we show that hydrogenated FNDs embedded in a polymer-based solid-state thin film device can be used for all-optical sensing of electric fields based on the NV charge state. The FND-doped polymer film is sandwiched in a capacitor device fabricated on a transparent conductive electrode, and an external AC voltage is applied to create a local electric field at the location of the nanodiamond. More than 95% of FNDs show an increase of up to 31% in $NV^-$ PL and a decrease of up to 13% in $NV^0$ PL upon application of an external voltage. We show that the change in NV PL increases monotonously with increasing voltage between 0 and 100 V, corresponding to an electric field from 0 to 625 kV cm$^{-1}$. The electric field sensitivity of a single FND and a single NV center is 19 V cm$^{-1}$ Hz$^{-1/2}$ and 72 V cm$^{-1}$ Hz$^{-1/2}$, respectively, assuming shot noise-limited detection. We investigate the NV charge state photodynamics on the millisecond timescale and find that the change in NV PL occurs within less than 0.1 ms after application of an external voltage and strongly depends on the laser excitation intensity and hence the rate of NV photoexcitation. We present a model that qualitatively accounts for the observed variations in NV PL by attributing them to an electric field-driven redistribution of photoexcited electrons from substitutional nitrogen defects to NV centers. This process results in a temporary conversion of $NV^0$ to $NV^-$ centers when an external voltage is applied. Our findings support the advancement of FNDs as all-optical, ratiometric sensors for electric fields and voltages in solid-state environments.

**Results**. Figure 1 a) shows a schematic illustration of the investigated device. An electrically insulating 250 nm thick layer of polyoctadiene (POD) was deposited via plasma polymerisation onto a commercial indium-tin-oxide (ITO)-coated glass substrate. 100 nm FNDs with a hydrogen surface termination (Adamas Nanotechnologies, USA) were sonicated and dispersed in a 1:3 mixture of water and 1-propanol at a concentration of 0.25 mg mL$^{-1}$. A ~400 nm thick FND-doped PVP thin film was spin-coated on the POD layer, followed by the deposition of a second insulating POD layer and a gold electrode. See Supporting Information (SI) Figures S1 and S2 for a photo of the device and polymer layer thickness measurements. An external voltage was applied via the ITO and gold electrode using a function generator and voltage amplifier. Transient charge spectroscopy, current-voltage (IV), and impedance measurements were used to verify that the resulting device was insulating and generated an electric field of 625 kV cm$^{-1}$ at 100 V applied voltage (see SI Figures S3 to S5). The device was investigated using a custom-built confocal PL setup. 532 nm laser excitation was delivered through the transparent conductive substrate using a 100× objective, and PL was collected with the same objective. See the Methods section and SI for more details on the device fabrication experimental setup. Figure 1 b) shows a confocal PL image of the x-y plane of the PVP layer. Individual FNDs or small aggregates of FNDs can be clearly identified throughout the PVP layer. A PL depth scan in the x-z plane perpendicular to the device layer structure (Figure 1 c) shows that the FND PL is strongly confined in the z-direction, in agreement with the ~400 nm thickness of the FND-doped PVP layer. Figure 1 d) shows a representative PL spectrum from an FND embedded in the PVP matrix exhibiting $NV^0$ and $NV^-$ zero-phonon lines at 575 nm and 637 nm, respectively (black trace), and pure $NV^0$ and $NV^-$ spectra (orange and red traces, respectively).

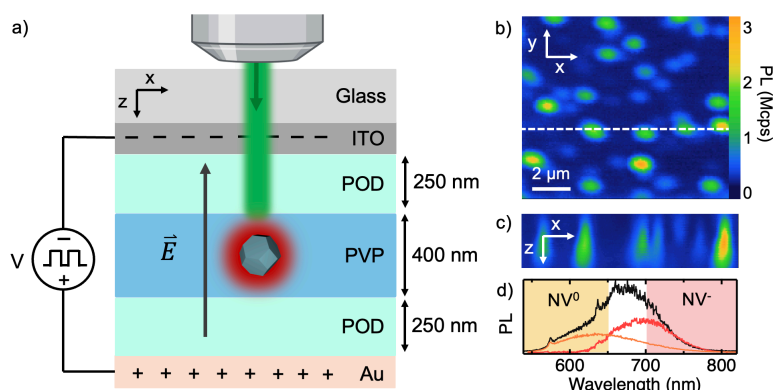

**Figure 1**. a) Schematic illustration of the multilayer capacitor device. The device was fabricated on an indium-tin-oxide (ITO)-coated glass substrate. It consists of two insulating dielectric layers (polyoctodiene, POD) that separate a fluorescent nanodiamond (FND) doped polymer (polyvinylpyrrolidone, PVP) layer from the ITO and gold electrodes. Square voltage pulses on the millisecond timescale create a local electric field $\vec{E}$ when the voltage was switched on ($V_{on}$= 100 V), and zero voltage ($V_{off}$) is applied between pulses. b) Confocal PL image of FNDs dispersed in the PVP layer imaged in the plane of the PVP layer (x-y-plane) acquired using 532 nm excitation and collecting PL above 550 nm. c) Confocal depth scan in the x-z-plane along the dashed line in panel b. d) Typical PL spectrum of an FND particle in the PVP layer (black trace) and pure $NV^-$ (red trace) and $NV^0$ spectra (orange trace). $NV^-$ and $NV^0$ PL signals were separated using optical long-pass (750 nm) and short-pass (650 nm) filters indicated by the red and orange shaded areas, respectively.

We first investigated the effect of applying an external voltage of +100 V on the PL intensity of many individual FNDs. In a typical experiment, the excitation beam (532 nm, 200 µW) was focused on an FND and the PL intensity was recorded with an avalanche photodiode over time with a time resolution of 0.1 ms. $NV^0$ and $NV^-$ charge states were recorded separately using 650 nm short-pass and 700 nm long-pass filters, respectively, as illustrated in Figure 1 d).

Figure 2 a) shows the change in $NV^-$ (red trace) and $NV^0$ (orange trace) PL over time as the external voltage is switched on for 15 ms at t = 0 and then switched off again. The data shown in Figure 2 a) was acquired for a total of 8 minutes with a 33% $V_{on}/V_{off}$ duty cycle ($V_{on}$ = +100 V for 15 ms, $V_{off}$ = 0 V for 30 ms) and averaged. Immediately after the voltage was switched on at t = 0 ms, the PL intensity collected in the $NV^-$ spectral band increased by 28% ($\Delta PL_{on}$), and the intensity collected in the $NV^0$ spectral band decreased by 13%. It is important to note that this change doesn't directly correspond to $NV^-$ and $NV^0$ PL intensities due to the strong spectral overlap of the two charge states, i.e. $NV^-$ PL contributes to the signal collected in the $NV^0$ spectral region and vice versa. Hence, in the case of an increase in $NV^-$ and decrease in $NV^0$ PL, $\Delta PL_{on}$ underestimates the actual change in both charge states' PL intensity. This will be discussed in more detail in the context of our proposed model below. Unless stated otherwise, in the following, the expressions '$NV^-$ PL' and '$NV^0$ PL' will refer to the PL intensities collected in the spectral windows >700 nm and 550-650 nm, respectively.

$NV^-$ and $NV^0$ PL signals decayed quickly within the first 5 ms and remained at about 2% above ($NV^-$) and 1% below ($NV^0$) the PL before the voltage was switched on until the voltage was switched off at t = 15 ms. When the voltage was switched off, both signals spiked again with a lower amplitude of +19% ($NV^-$, $\Delta PL_{off}$) and -7.1% ($NV^0$) and returned to close to the PL intensity before the voltage was applied within less than 10 ms. Figure 2 a) shows the $NV^-$ and $NV^0$ signals acquired from one FND in the PVP layer. To test the reproducibility of the observed signals, we investigated the same particle multiple times over the course of a day and found no significant change in the amplitude and dynamics of the signals.

We then investigated if the changes in NV PL shown in Figure 2 a) are the typical response of FNDs in the capacitor device upon application of an external voltage. We investigated the effect of an applied voltage ($V_{on}$ = +100 V) on the $NV^-$ PL intensity for 110 particles. The signal from 106 particles (96.4%) shows two pronounced positive peaks when the voltage was switched on and off, as shown in the red trace in Figure 2 a), while the amplitude of the peaks varies. We find that the change in $NV^-$ PL is always positive, while that of $NV^0$ is always negative. Figures 2 b) and c) show histograms of the maximum change in PL for the first peak ($\Delta PL_{on}$) and the ratio between the first and the second peak amplitude ($\Delta PL_{on}/\Delta PL_{off}$), respectively. While 52% of the particles showed a $\Delta PL_{on}$ between 2% and 12%, 7% of particles showed a peak increase in $NV^-$ PL intensity above 20%. The $\Delta PL_{on}/\Delta PL_{off}$ ratio shows an approximately Gaussian distribution centered around 1 with a FWHM of 0.45. Figure 2 d) shows a histogram of the steady state change in PL after the voltage was switched on and the peak decayed ($\Delta PL_{ss}$). More than 90% of particles showed a $\Delta PL_{ss}$ between -2% and +2%, with a third of particles exhibiting no steady state change in PL ($\Delta PL_{ss}$ = 0 ± 0.25%). See SI Figure S6 and S7 for examples of the different types of responses observed.

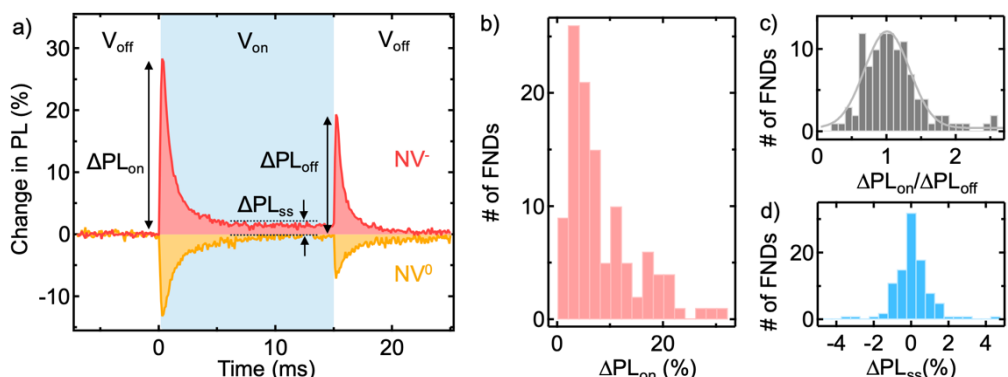

**Figure 2.** The effect of an external voltage on the PL of FNDs embedded in a PVP thin film capacitor. a) Change in PL $NV^-$ (red trace) and $NV^0$ (orange trace) PL as a function of time. An external voltage of $V_{on}$=+100 V is applied from time t=0 to 15 ms, and no voltage is applied at all other times ($V_{off}$ = 0 V). b) Histogram of the change in NV⁻ PL immediately after the voltage was switched on ($\Delta PL_{on}$) c) Histogram of the ratio between the change in PL when the voltage was switched on and switched off ($\Delta PL_{on}/\Delta PL_{off}$). d) Histogram of the steady state change in PL ($\Delta PL_{ss}$) after the voltage was switched on and the peak has decayed. The histograms are based on measurements for 110 FNDs.

We then investigated the effect of the amplitude of the applied voltage on FND PL. Figure 3a) shows the change in NV⁻ PL over time for 40, 60, and 100 V applied at time t = 0 ms. In all cases, the NV⁻ PL increased immediately when the voltage was applied and decayed within 2-4 ms to the steady state PL ($\Delta PL_{ss}$). $\Delta PL_{on}$ increased with increasing voltage from a 6.0% increase for 40 V to a 26% increase for 100 V. The decay time of each peak, however, is not affected by the applied voltage (see SI Figure S8).

Figure 3 b) shows $\Delta PL_{on}$ as a function of applied voltage for the particle investigated in Figure 3 a), as well as a second FND. In both cases, $\Delta PL_1$ increased monotonically with increasing applied voltage. While the PL from both FNDs only increased by 1-2% when the voltage increased from 0-20 V, the PL increased by 8.5% (FND1) and (3.7%) for a voltage increase from 60 V to 80 V. Towards 100 V, the PL from FND1 shows saturation while the PL from FND2 does not. Voltages above 100 V were not explored to avoid dielectric breakdown. For particles with $\Delta PL_{ss}$ values above 0, we find that $\Delta PL_{ss}$ also increased with increasing applied voltage (see SI Figure S9).

For a bright 100 nm FND, we typically detect ~ 3 M photocounts per second in total (Figure 1 b), ~25% (0.75 M) of which are emitted >700 nm where we collect NV⁻ PL. From the slope of FND1 in Figure 3 b) between 40 and 80 V and assuming shot-noise-limited collection of 0.75 M cps, we determine an electric field sensitivity of 19 V cm⁻¹ Hz⁻¹ᐟ² for the FNDs in our devices. For PL from a single NV in an optimized optical system[18] (200k cps, shot-noise-limited) using the same slope and spectral filtering, the estimated electric field sensitivity is 72 V cm⁻¹ Hz⁻¹ᐟ², which is more than an order of magnitude improvement over the ODMR-based electric-field sensitivity of 891 V cm⁻¹ Hz⁻¹ᐟ² reported for single NVs in bulk diamond.[3] See SI Figure S12 for details.

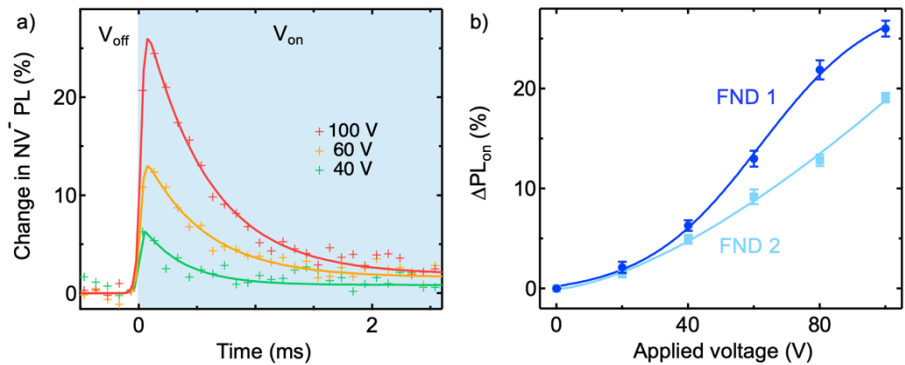

**Figure 3.** a) Change in NV⁻ PL as a function of time for various applied voltages. Markers show the experimental data, and solid traces are fits using a step function multiplied by a stretched exponential. The voltage was applied at time t = 0 ms, as indicated by the blue shaded area. b) Maximum change in NV⁻ PL ($\Delta PL_{on}$) as a function of applied voltage for two FNDs. Markers are the experimental data and error bars represent the error in the fit to the experimental data. Solid traces are a guide to the eye only.

The laser excitation intensity strongly influences the NV charge state equilibrium and the underlying charge state cycling processes.[19] Hence, we studied the effect of the 532 nm laser excitation power ($P_{ex}$) on the dynamics and amplitude of the voltage-induced PL change. Figure 4 a) shows the normalized change in NV PL intensity (NV⁻ and NV⁰) as a function of time after an external voltage of 100 V was applied at t = 0 ms for excitation intensities from 30 to 200 μW. The solid lines show a stretched exponential fit (stretch exponent 0.7) to the experimental data (markers) and reveal a significant increase in the decay rate with increasing excitation intensity. The change in PL decreased by a factor of 10 within 3.8 ms for the lowest excitation intensity (30 μW), while the same decrease occurred within 0.3 ms for 200 μW. Figure 4 b) shows the decay time τ determined via the stretched exponential fits in Figure 4 a) as a function of excitation intensity. The solid black line is an exponential fit to the τ values. τ rapidly decreases with increasing excitation power and approaches the time resolution of the measurement (0.1 ms) at an excitation power of 200 μW.

Having established that the dynamics underlying the voltage-induced PL changes become faster with increasing excitation intensity, we then investigated the effect of the excitation intensity on the amplitude of the change in NV⁻ and NV⁰ PL. Figure 4 c) shows the maximum change in PL when the voltage was switched on ($\Delta PL_{on}$) as a function of laser excitation power for NV⁻ (red markers and trace) and NV⁰ (orange markers and trace). The change in NV⁻ PL increased from 12% to 28% as the excitation intensity increased from 30 to 60 μW, then remained constant within a 2% margin and then dropped above 200 μW to 9% at 900 μW excitation intensity. The change in NV⁰ PL mirrors this behaviour with negative changes of lower absolute value. The change in PL from NV⁻ and NV⁰ is maximized for excitation intensities between 60 and 200 μW. The decrease at higher $P_{ex}$ values may be explained by faster NV charge state cycling at higher excitation intensities or by the decay becoming much faster than our time resolution of 0.1 ms, leading to an apparent decrease in signal intensity. This will be discussed in more detail in the context of our proposed model.

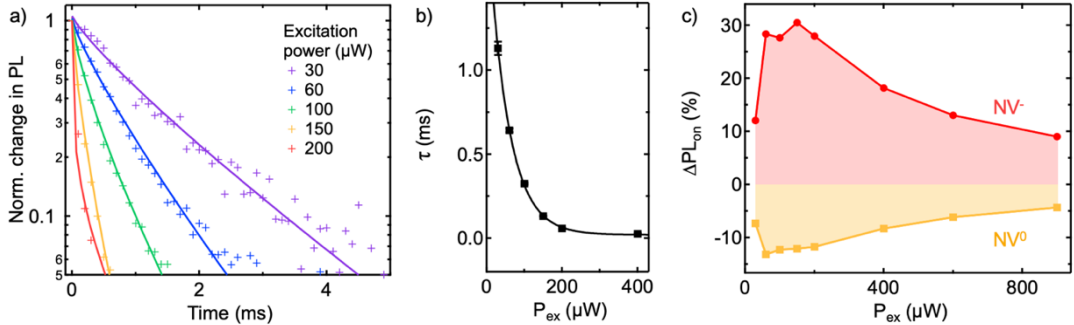

**Figure 4**. The effect of the laser excitation power on NV PL dynamics. a) Normalized change in NV PL as a function of time for different excitation powers. Markers show the experimental data and solid traces are stretched exponential fits to the data. b) PL decay time τ as a function of laser excitation power $P_{ex}$. The solid black line is an exponential fit to the τ values (markers). c) Maximum change in PL when the voltage was switched on (ΔPL$_{on}$) vs laser excitation power for NV⁻ (red markers and trace) and NV⁰ (orange markers and trace).

We propose that our observations can be explained by an electric field-induced charge state modulation of NV centers in FNDs. Figure 5 a) shows an illustration of the change in NV charge state population as the voltage is switched on. NV⁻ and NV⁰ populations are constant before the voltage is switched on. Application of an external voltage leads to a strong transient increase and decrease in NV⁻ and NV⁰ populations, respectively, followed by a steady state, where both populations return to levels slightly above (NV⁻) and below (NV⁰) those observed before the voltage is applied. On average, our experiments show a 48 ± 12 % lower absolute value of ΔPL$_{on}$ for NV⁰ than for NV⁻. This difference may be explained by a combination of (1) the spectral windows where NV⁻ and NV⁰ PL is collected, and (2) a conversion of NV⁰ to non-fluorescent NV⁺. (1) Due to the strong spectral overlap of NV⁰ and NV⁻, we expect a 10% increase and decrease in NV⁻ and NV⁰ populations, respectively, to lead to a PL increase of 8.0% for NV⁻ (> 700 nm) and a 3.9% PL decrease for NV⁰ (550-650 nm) in the spectral regions we investigate (see Figure S10). This effect alone would suggest that the absolute value of the change in PL of NV⁰ is only 49% of the change in PL of NV⁻ in the case of a one-to-one conversion of NV⁻ to NV⁰ in our measurements, and could explain the observed 48 ± 12 % smaller NV⁰ signal compared to the NV⁻ signal. However, this difference in NV⁰ and NV⁻ signal amplitude strongly depends on the initial charge state ratio in the absence of an external field, which is known to vary between individual particles.[20] (2) The total PL collected from both charge states is not necessarily conserved due to the creation of NV⁺, which is non-fluorescent. The creation of NV⁺ has been reported to occur at a hydrogenated bulk diamond surface upon application of an external voltage[5] and is likely to occur for NV centers near the FND surface in our experiments.

Before an external voltage is applied and under constant-intensity light excitation, NV charge state cycling takes place on the microsecond timescale,[21,22] and time-averaged NV⁻ and NV⁰ populations remain constant on the millisecond timescale in our experiments. Near the hydrogenated FND surface, most NVs are in the NV⁰ charge state due to surface-induced band bending (Figures 5 b and d). Figure 5 e) illustrates the near-surface band bending we calculated based on a 1D model of a hydrogenated FND (see SI Figure S11).

Upon application of the external voltage, the local electric field leads to a redistribution of charges inside and around the FND (Figures 5 c and f). Neutrally charged substitutional nitrogen ($N_s^0$) is generally present at concentrations on the order of 100 ppm in the high-pressure high-temperature FNDs investigated here, which is at least ten times higher than the NV concentration. $N_s^0$ is the main electron donor for NV centers ($N_s^0$ → $N_s^+$ + e⁻) and photoexcited electrons from $N_s^0$ are known to strongly contribute to NV charge state cycling under excitation with light below 560 nm (>2.2 eV)[23,24]. As charges rearrange inside the FND upon application of the external voltage, electrons photoexcited from $N_s^0$ are transiently trapped by nearby NV centers, leading to an increase in NV⁻ and decrease in NV⁰ populations (Figure 5 f). This hypothesis is supported by the strong light excitation intensity dependence of the PL dynamics shown in Figure 4. At low excitation intensities, fewer photoexcited electrons are available, leading to a smaller ΔPL$_{on}$ amplitude (Figure 4 c). With increasing excitation intensity, more photoexcited electrons are created and transiently trapped by NV centers, leading to an increase in the |ΔPL$_{on}$| amplitude. At the same time, more charge carriers are available per unit time to create a local screening field ($E_s$, Figure 5 d), causing a faster decay in the observed change in PL (Figure 4 a). At the highest excitation intensities (>300 µW), these processes become faster than the temporal resolution of our measurements (0.1 ms), leading to an apparent decrease in the observed |ΔPL$_{on}$| amplitude. A strong increase in NV charge state cycling rates at high excitation intensities could make charge state cycling the dominating process in this regime and thereby also contribute to the latter observation. In principle, the above could also apply to mobile positive charge carriers, but the strong increase in NV$^-$ suggests that electrons dominate the observed photodynamics – potentially due to the large excess of N$_s^0$ electron donors in our particles.

Once redistributed, the charges induce an electric field $E_s$ (Figure 5 d) that screens the field induced by the applied voltage, reducing the ΔPL absolute values for NV$^-$ and NV$^0$ to their steady state values, ΔPL$_{SS}$. While the distribution of ΔPL$_{SS}$ values observed for 110 FNDs is centred around zero (Figure 2d), most particles exhibit some degree of PL modulation in a constant E-field, which increases with increasing applied voltage (SI Figure S9). The source of ΔPL$_{SS}$ can be explained by the near-surface upward band bending predicted by the equilibrium model of charge distribution in the FNDs, as illustrated in Figure 5d. Under a constant E-field, Figure 5e, the equilibrium reached after charge redistribution results in an asymmetric modulation in NV charge states. This effect can be replicated by solving Poisson's equation for a simple 1D model of the FNDs, as shown in SI Figure S11. In our model, the sign of ΔPL$_{SS}$ depends on the degree of the FND surface hydrogen termination. In experiments, the sign of ΔPL$_{SS}$ for a given FND likely also depends on other factors, including the particle shape, its orientation in the external field, and interaction with the polymer matrix. When the external voltage is switched off, the screening field acts analogously to the external electric field and leads to the transient redistribution of photoexcited electrons described above, causing the second spike in NV$^-$ and NV$^0$ PL.

Different types of local charges may be involved in the charge redistribution process, including FND surface charges, charges associated with lattice defects (including N$_s^0$ as discussed above), and local polymer charges. Based on our results, we are unable to determine which charges are mainly responsible for the creation of the screening field $E_S$. However, our results do show that photoexcited electrons inside the FNDs are a major contributor to the observed electric field-induced changes in NV PL.

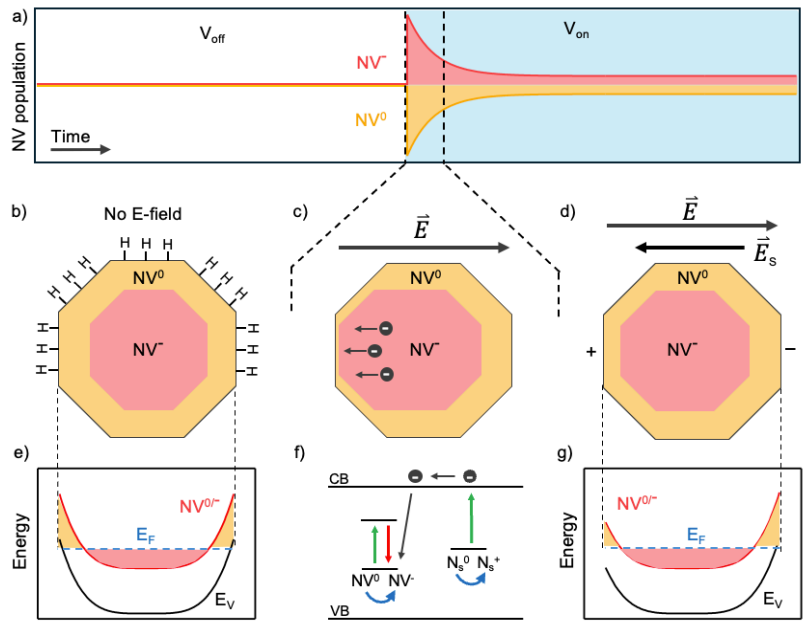

**Figure 5.** Proposed mechanism underlying the observed voltage-induced PL changes. a) Illustration of the change in NV$^-$ and NV$^0$ populations as a function of time as the voltage is switched on. b) to d) Schematic illustrations of a hydrogenated FND, the spatial distribution of the different NV charge states, and the movement of photoexcited electrons before (b) and after (c and d) the voltage is applied. e) & g) Band diagrams showing the diamond valence band (E$_V$), Fermi level (E$_F$) and the NV$^{0/-}$ transition energy across an FND in the absence (e) and in the presence of an external electric field once a steady state has been reached (g). f) Energy diagram of NV and substitutional nitrogen (N$_s$) illustrating the transient electric field-induced redistribution of photoexcited electrons from N$_s$ to NV defects in the first 3 ms after the voltage is switched on.

Our study is the first to investigate FNDs for electric-field sensing in a solid-state capacitor device. Karaveli et al showed that the charge state of NVs in FNDs in an aqueous electrolyte-based electrochemical cell can be modulated by an external voltage[10] and our results qualitatively agree with the observations in this study. However, a direct comparison of results is challenging since FNDs in [10] were in direct contact with a conductive electrolyte solution, a conductive substrate, or both, and an electric field sensitivity was not reported. Charge state-switching of near-surface NV centers in bulk diamond in contact with an electrolyte solution was also reported by Gotz et al.[4]

In bulk diamond, an electric field sensitivity of 891 V cm$^{-1}$ Hz$^{-½}$ was reported for single near-surface NV centers based on the electric-field-induced shift of the ODMR resonance frequency,[3] which is an order of magnitude higher than the single NV sensitivity of 72 V cm$^{-1}$ Hz$^{-½}$ reported here. ODMR-based electric-field sensing using NV centers in FNDs has not been reported. The ODMR-based electric field sensitivity of NVs in FNDs is expected to be significantly higher compared to bulk due to the broadening of ODMR lines in FNDs compared to bulk.[25] Hence, we expect NV charge state-based sensing to outperform ODMR-based approaches in FNDs by more than one order of magnitude.

Nanoscale electric field imaging was reported using ODMR measurements of near-surface NV centers in bulk diamond in combination with a charged atomic force microscopy tip[8] with a significantly higher sensitivity of 35.2 kV cm$^{-1}$ Hz$^{-½}$. The same study reported reliable control of the NV charge state by controlling the local concentration of positive and negative charges near the NV center at the diamond surface. Under green light excitation, the electric field-driven accumulation of photoexcited electrons at the surface was proposed as the mechanism underlying charge state switching to NV$^-$.

Future studies will focus on optimising charge-state-based sensing in solid-state devices. Ratiometric detection can be implemented by simultaneously collecting PL in NV$^0$ and NV$^-$ spectral bands, which will make the PL readout more robust by removing intrinsic fluctuations in the excitation and emission intensities. The width and center wavelength of the spectral bands can be optimised to reduce the overlap of NV$^0$ and NV$^-$ signals in the two spectral channels. Different FND particle sizes and surface chemistries can be investigated to maximise the PL signal amplitude as well as the excitation wavelength to control the NV charge state cycling. Optimisation of these parameters is expected to enable significant improvements in the voltage sensitivity of FNDs in solid-state systems.

In summary, we show that hydrogenated FNDs embedded in a polymer-based capacitor device can be used for the all-optical detection of applied voltages based on the NV charge state. We show that 95% of FNDs show a transient increase in NV$^-$ PL of up to 31% within 0.1 ms after an external voltage is applied, accompanied by a transient decrease in NV$^0$ PL (Figure 2 a). The change in PL increases with increasing applied voltage from 0 to 100 V (Figure 2) and we infer an electric field sensitivity of 19 V cm$^{-1}$ Hz$^{-½}$. We find that the PL dynamics in the first 5 ms after the voltage is applied (Figures 4a and 4b) and the maximum PL change (Figure 4c) strongly depend on the light excitation intensity. Based on this observation, we propose that photoexcited electrons from $N_s^0$ redistribute inside FNDs (Figure 5 c) as a result of the local electric-field induced band bending, and are transiently trapped by NV$^0$ centers to create NV$^-$ (Figure 5 f) leading to the observed increase in NV$^-$ and decrease in NV$^0$ PL. The redistributed charges inside and around FNDs create a screening field opposing the external electric field. Once a steady state is reached, the net local electric field leads to an asymmetric modulation of the NV$^-$/NV$^0$ transition levels near the FND surface (Figure 5d & 5g). Depending on the degree of surface hydrogen termination, this can lead to an increase or decrease in the change in NV$^-$ PL ($\Delta PL_{SS}$) in agreement with our observations (Figure 2d). Our findings will contribute to the development of FNDs embedded in solid-state devices as robust, all-optical voltage sensors based on the NV charge state.

## Methods
### Capacitor fabrication
The capacitor was constructed upon an indium-tin-oxide (ITO) coated glass substrate (18 ×18 mm) (SPI supplies, USA) with a resistance of 30 –60 Ω cm$^{-2}$. 1-7 octadiene was deposited to a thickness of 250 nm using plasma polymerisation, forming poly-octadiene (POD) (10$^{-4}$ mbar chamber pressure, 2.25 N mL min$^{-1}$ flow rate, 50 W). The same procedure was repeated after the deposition of the FND-doped polymer layer. FND-doped polymer thin films: polyvinylpyrrolidone (PVP, MW 40,000) was dissolved in 1-propanol at a concentration of 5% wt/vol. Commercially available (Adamas Nanotechnologies, USA) 100 nm hydrogenated nanodiamonds in water (1 mg mL$^{-1}$) were sonicated for 10 minutes, then combined with the PVP solution and additional DI water, resulting in an FND concentration of 0.25 mg mL$^{-1}$ and a 1:3 ratio of water and 1-propanol. 100 µl of the PVP-FND solution was then spin-coated onto ITO-coated glass and poly-octadiene. The solution was statically dispensed before spin coating at 1800 RPM with an acceleration of 2000 RPM/s, resulting in a PVP thickness of ~400 nm. The second POD layer was then deposited, followed by chromium gold sputtered electrodes to a thickness of 50 nm.

**Optical measurements and voltage delivery**
All optical measurements were performed using a custom-built confocal fluorescence microscope. NV centres were excited with a 532 nm continuous-wave laser (Opus 2 W, Laser Quantum, Lastek, Australia) through a 100× objective (TU Plan Apo, NA = 0.90, Nikon, Japan). Fluorescence was separated from the excitation path using a 538 nm dichroic beamsplitter (Semrock, USA), followed by spectral filtering: a 550 nm long-pass and 650 nm short-pass filter for $NV^0$ detection, or a 700 nm long-pass and 900 nm short-pass filter for $NV^-$ detection (all from Thorlabs, USA). Photons were detected with an avalanche photodiode (Excelitas, USA). Time trace acquisition and electric field control were implemented via custom LabVIEW software. Photons were counted in 0.1 ms bins (10 kHz) over 45 ms, resulting in 4500 data points per trace. Each trace included a 15 ms zero-field period, followed by a 15 ms, 100 V field pulse, and another 15 ms zero-field interval (see SI Figure S1). This sequence was repeated and averaged over up to 6000 iterations. A data acquisition board (NI USB 6001) connected to a 10x operational amplifier (PDu150, Piezo Drive, Australia) was used as a function generator to apply voltage to the capacitor via a custom printed circuit board (PCB) (see SI Figure S1). The capacitor was glued over an optical access hole in the PCB and the top and bottom electrodes wire-bonded to contact pads.

**1D Poisson model**
Nanodiamonds were model as 100 nm thin sheets in a 1D approximation. Each of the two surfaces had an areal density of $sp^2 = 5 \times 10^{12}$ $cm^2$ defects and hydrogen termination induced adsorbed surface acceptors with an areal densities of $Q_{SA} = 0$ to $2 \times 10^{13}$ $cm^2$. The bulk diamond had 10 ppm $N_s$ defects and 0.1 ppm NV defects. The potential bands within the diamond model were solved using the python *diamond-bandalyzer* package (available on PyPi), which uses a Newton-Rhaphson minimisation routine. To estimate the potential bands within an FND that had redistributed its charges to screen a constant E-field, an unvarying linear background potential was added to the band potential during minimisation, ensuring a potential drop of $V_{BG} = 1$ V occurred across the 100 nm of diamond. The Fermi level of the final system was calculated assuming the diamond remained charge neutral.

**Acknowledgements**
PR acknowledges support through an Australian Research Council (ARC) DECRA Fellowship (DE200100279), an ARC Discovery Project (DP250100125), and an RMIT University Vice-Chancellor's Senior Research Fellowship.

# All-optical electric field sensing with nanodiamond-doped polymer thin films

Roy Styles[1*], Mengke Han[2], Toon Goris[3], James Partridge[1], Brett C. Johnson[1], Blanca del Rosal[1], Amanda N. Abraham[1], Heike Ebendorff-Heidepriem[2], Brant C. Gibson[1], Nikolai Dontschuk[3], Jean-Philippe Tetienne[1*], Philipp Reineck[1*]

[1] School of Science, RMIT University, VIC, 3001
[2] University of Adelaide, Adelaide, SA 5005
[3] School of Physics, University of Melbourne, Parkville, Victoria 3010, Australia

* roy.styles@rmit.edu.au, jean-philippe.tetienne@rmit.edu.au, philipp.reineck@rmit.edu.au

## Device circuit and time trace acquisition

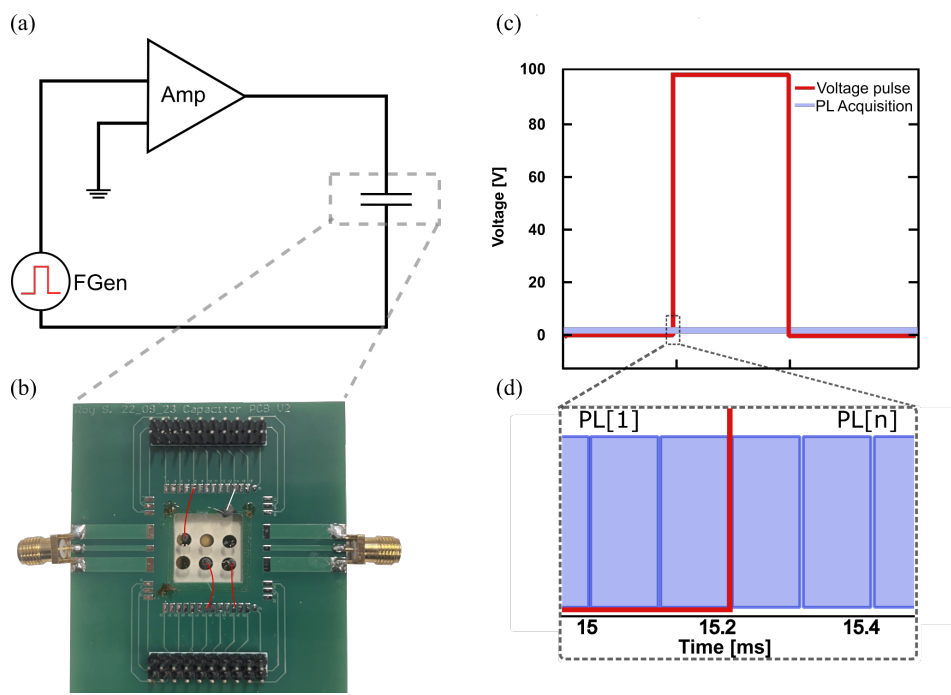

**Figure S1:** **(a)** Schematic of the circuit used to apply a voltage across the multilayer capacitor device, consisting of a function generator, an operational amplifier, and the capacitor device. **(b)** Photograph of the custom-printed circuit board (PCB) designed for connection and measurement of the multilayer capacitor, which is glued at the center of the chip (glass square with six circular metal electrodes). The capacitor is mounted over a central hole to allow optical access through the transparent indium tin oxide (ITO) top layer (see Figure 1, main text), while keeping the capacitor electrodes accessible for wire bonding. Wire bonds in the photograph to the gold electrodes and ITO layer are highlighted with red and white lines, respectively. The gold electrodes are connected to rectangular bonding pads on the PCB, which are routed to two rows of header pins for integration with the rest of the circuit. The PCB also includes BNC connectors soldered to printed strip lines for microwave signal application if needed. The system is controlled by custom software built in LabVIEW that outputs a trigger signal that initiates APD PL acquisition as well as the function generator. Each triggered APD acquisition is represented by the blue rectangles in **(d)** while the function generator output is shown in **(c)** by the red square wave. Within the control software the acquisition rate can be customized as well as the size and duration of the function generator output. In the measurements presented in the main text, the acquisition rate is 10 kHz, while the function generator is triggered at a rate of ~33.3 Hz. The square wave from the function generator is amplified 10x by the operational amplifier.

## Derivation of theoretical capacitance

The voltage drop and capacitance of the multilayer dielectric capacitor can be approximated by three separate capacitors in series [1]. Therefore, we can use the following derivation to determine the capacitance and voltage drop across individual layers. For capacitors in series, the charge $Q$ on each capacitor is equal, while the total voltage is the sum of voltage drops across each layer. The voltage drop across an individual capacitor is:

$$V_i = \frac{Q}{C_i} \tag{1}$$

Summing over all three layers, the total voltage across the multilayer structure is:

$$V_{total} = \sum_{i=1}^{3} V_i = Q\left(\frac{1}{C_1} + \frac{1}{C_2} + \frac{1}{C_3}\right) \tag{2}$$

Therefore, the total capacitance across the structure may be written as:

$$\frac{1}{C_{total}} = \frac{1}{C_1} + \frac{1}{C_2} + \frac{1}{C_3} \tag{3}$$

Each individual layers capacitance $C_i$ is given by the parallel-plate capacitance formula:

$$C_i = \frac{\epsilon_0 K A}{d_i} \tag{4}$$

Where:

- $\epsilon_0$ is the vacuum permittivity (8.85418782 × 10$^{-12}$ F/m).
- $A$ is the area of the capacitor electrode ($\pi 0.0015^2$).
- $d_i$ is the thickness of the dielectric layer (416 nm for PVP, 250 nm for POD).
- $K$ is the dielectric constant of the material (7.7 for PVP, 3.21 for 1,7-POD).

Substituting Eq. (4) into Eq. (3), we obtain the total capacitance of the device in terms of its physical and material properties.

$$\frac{1}{C_T} = \left(\frac{d_1}{e_0 K_1 A} + \frac{d_2}{e_0 K_2 A} + \frac{d_3}{e_0 K_3 A}\right) \tag{5}$$

$$C_T = \frac{e_0 A}{\frac{d_1}{K_1} + \frac{d_2}{K_2} + \frac{d_3}{K_3}} \tag{6}$$

To determine the voltage drop across a particular layer $i$, we combine Eq. (1) with the expression fo

$$Q = C_{Total} V_{Total}$$

$$V_i = V_{Total} \frac{C_{Total}}{C_i} \tag{7}$$

or with the substitution of Eq. 6

$$V_i = \frac{V_{Total}}{C_i} \left(\frac{e_0 A}{\frac{d_1}{K_1} + \frac{d_2}{K_2} + \frac{d_3}{K_3}}\right) \tag{8}$$

This relation allows us to estimate the electric field experienced by individual layers. Equation 6 yields a total device capacitance of 0.301 nF which is in close agreement with the capacitance measurements acquired in Figure S11 at ~67 Hz.

## Layer thickness measurements

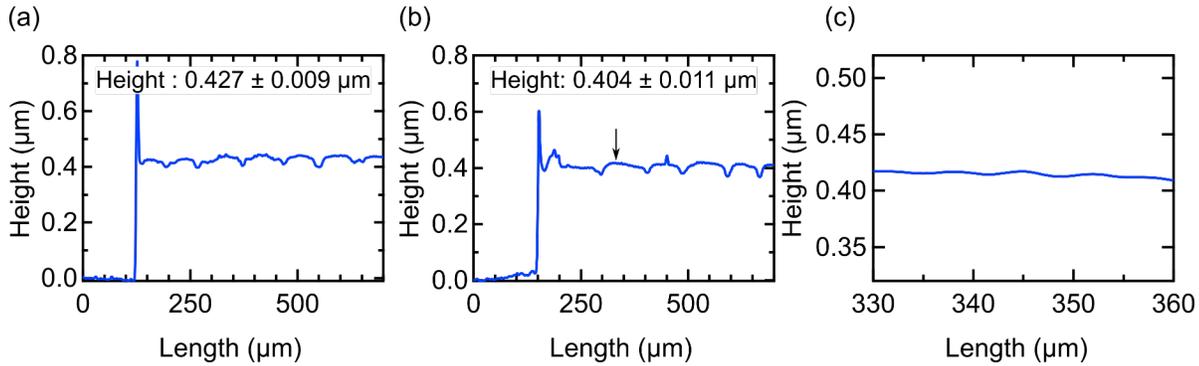

**Figure S2:** 2D stylus profilometer scans of PVP layers spin-coated on ITO-coated glass coverslips from two different regions (a) and (b). An arrow is included in (b) to indicate the approximate location of the zoomed in data shown in (c), included to highlight the smoothness of the layers over the width of a typical confocal scan (30 μm). The step at the beginning of each scan corresponds to the height difference between the bare glass substrate and the polymer-coated region. This step is created by removing a portion of the PVP layer using a cotton bud dipped in isopropyl alcohol (IPA) and a sharpened toothpick to define a clean edge. The exposed glass region is used as a height reference (zero point) during profilometry. These measurements are routinely performed during fabrication to optimize spin coating parameters for film uniformity and thickness. The film must be thick enough to fully encapsulate 100

nm nanodiamonds, while maintaining low surface roughness. These layers were spin-coated at 1800 RPM with an acceleration of 2000 RPM/s and a static dispense. This recipe yields an average film thickness of 0.416 µm and an average surface roughness of 0.010 µm. Sharp peaks observed at the left edge of each scan arise from polymer build-up at the wiped step edge.

## IV characteristics

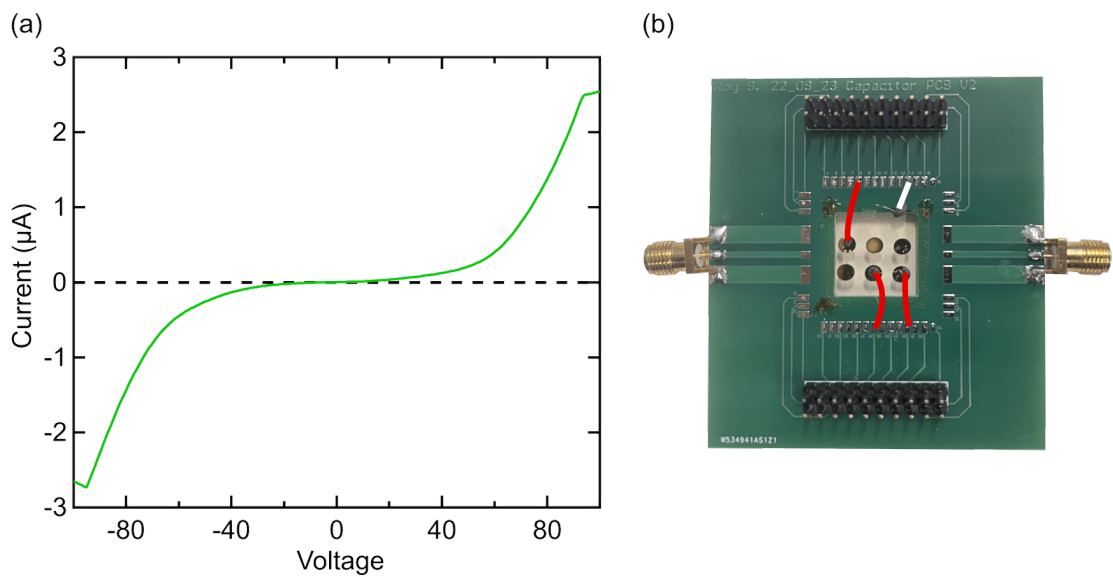

**Figure S3: (a)** Current–voltage (IV) characteristics of the capacitor, measured using a home-built IV acquisition program connected via a National Instruments data acquisition board (NI DAQ) and a trans-impedance amplifier (TIA). The current increases exponentially with applied voltage, reaching 2.7 µA at 100 V (equivalent to a resistance of 40 MΩ). At the minimum voltage used during PL measurements (20 V), the leakage current is 39 nA, corresponding to a resistance of 513 MΩ. These values are consistent with typical leakage levels observed in commercial capacitors and are negligible in the context of the model discussed in the main text. **(b)** Image of the device included to highlight negative and positive polarity. The positive connections are wire bonds highlighted in red bonded to the circular electrodes (bottom in Figure 1 of the main text). The negative connection is the wire bond highlighted in white bonded to the ITO layer (top in Figure 1 of the main text).

# Charge transience characterization

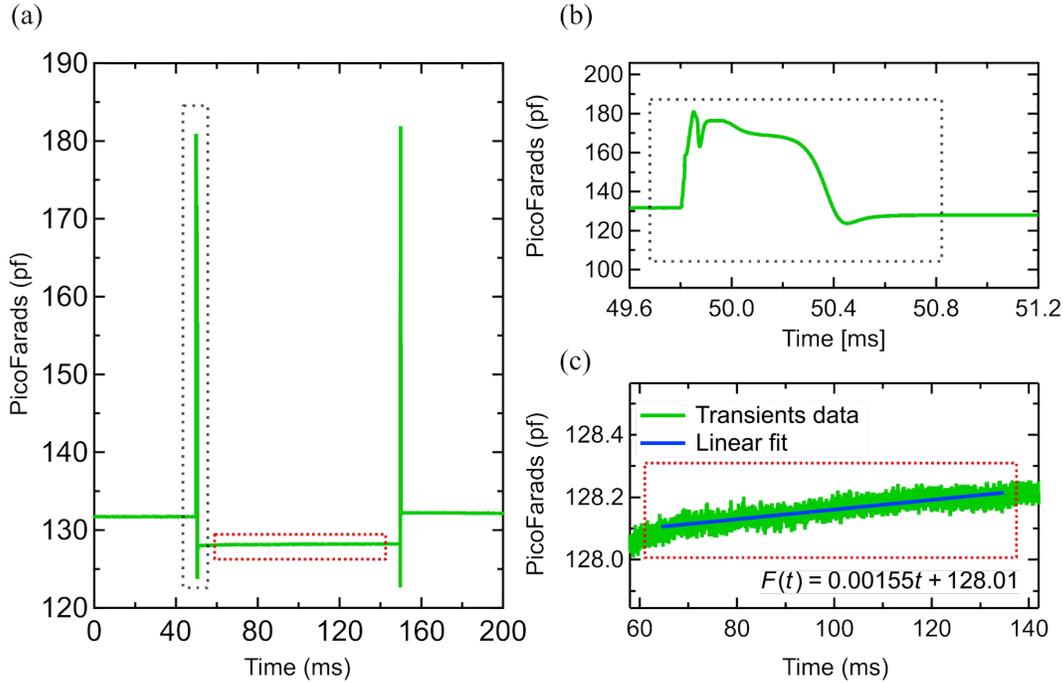

**Figure S4:** Charge transience characterization of the multilayer capacitor using deep-level transient spectroscopy (DLTS) performed with a Boonton 7200 capacitance meter. These measurements aim to assess the degree of charge movement within the device while applying a +100 V bias voltage to evaluate whether this charge movement may contribute to the observed photoluminescence (PL) response of the FNDs described in the main text. **(a)** Shows the entire measurement period. 100 V is applied across the capacitor at 50 ms and turned off at 150 ms, indicated by the change in capacitance. The decrease in capacitance of 3.54 pF (2.69%) when the voltage is applied indicates a small degree of initial charge movement throughout the capacitor. **(b)** Is a rescaled plot of the grey dotted area in **(a)**. The Boonton circuit introduces a ringing artifact due to its limitations in handling fast and high-amplitude signals (>20 V/ms), producing sharp peaks at voltage transitions (50 ms and 150 ms) [3]. These artifacts obscure the initial transient response, indicating that any charge motion occurs on the sub-millisecond timescale. This also suggests that the small global charge movement within the capacitor is not responsible for the observed PL changes, which take place on the 10 ms timescale. **(c)** Zoomed view of the red dotted region in **(a)**, here the capacitance is increasing gradually over the voltage-on (100 V) period. This is indicative of slight charge movement within the capacitor; however, it's impossible to determine their origin from this measurement. The linear fit (blue line) shows a capacitance increase of 0.00155 pF/ms over 70 ms, corresponding to a total capacitance change of 0.109 pF (0.085%) or an estimated charge movement of 0.109 pC.

## Impedance spectroscopy

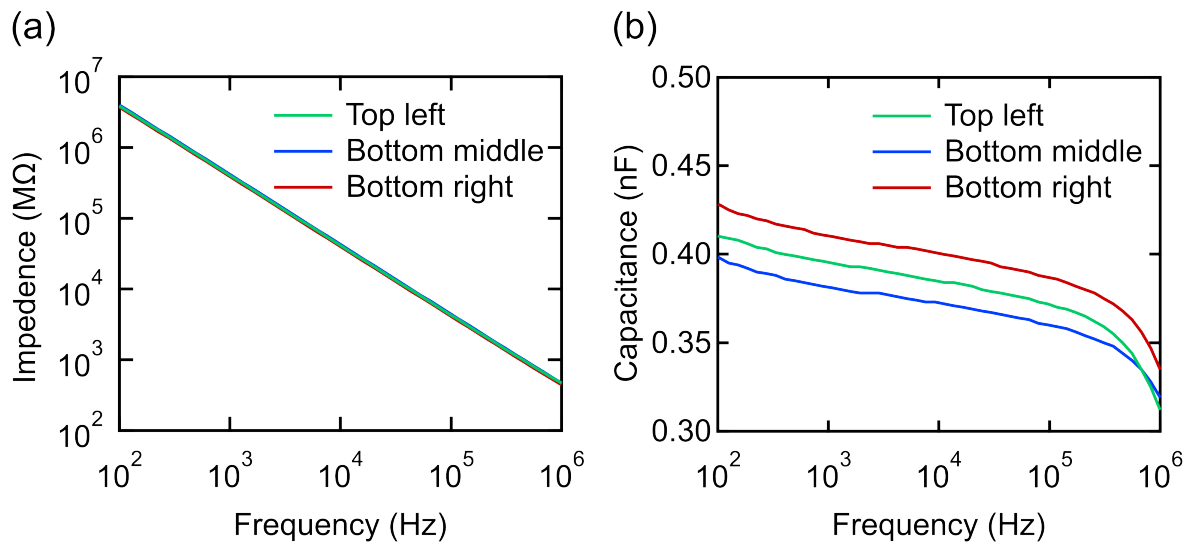

**Figure S5**: Impedance and capacitance measurements acquired from three electrodes on the capacitor (See Figure S1) using a Zurich Instruments MFIA digital impedance analyzer at 3 V. Only three electrodes were measured as they were the only electrodes successfully wire bonded during fabrication. All main-text measurements were performed using the "Top left" electrode (green trace), but additional electrodes were characterized in case of device failure. **(a)** Impedance as a function of AC frequency. The inverse dependence is characteristic of an ideal capacitor: at higher frequencies, there is less time for charge accumulation before polarity reverses, resulting in decreased opposition to current. The absence of flattening at low frequencies indicates minimal leakage current, confirming that the device performs as a near-ideal capacitor at low voltage. **(b)** Capacitance as a function of frequency. The operational frequency used in the main text (~67 Hz) yielded a measured capacitance of 0.416 nF, which is in close agreement with the theoretical value of 0.301 nF (see derivation above), supporting the conclusion that the fabricated capacitor exhibits near-ideal behavior.

## NV⁻ photoluminescence vs time for individual FNDs

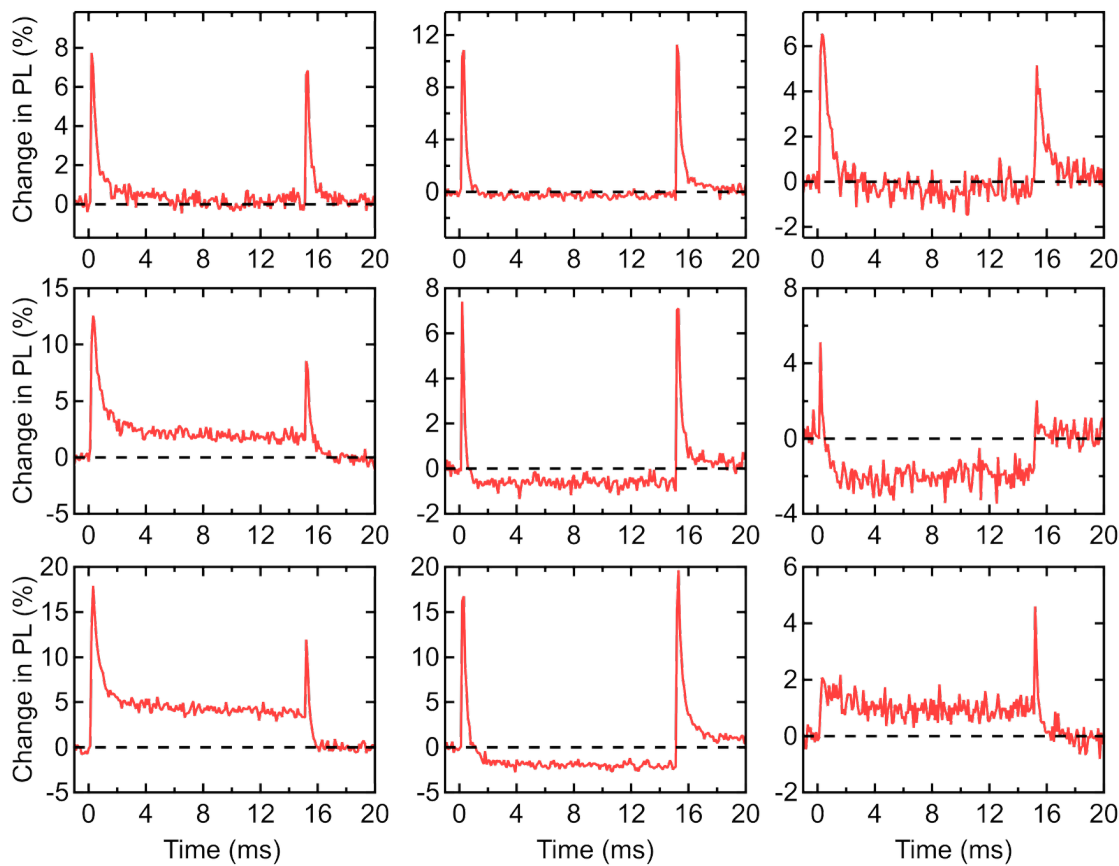

**Figure S6:** Change in NV⁻ PL intensity as a function of time for six different nanodiamonds as an external electric field (100 V) is switched on (t=0 ms) and off (t=15 ms). These time traces were acquired using 532 nm CW excitation (~150 µW) and PL collected in a spectral window of 700 – 850 nm to focus on the change in the NV⁻ charge state. The measurements were acquired as part of the statistical analysis of 110 particles, as described in Figure 2 in the main text, and show the different types of responses we observed across the investigated particles.

**NV⁻ and NV⁰ photoluminescence vs time for individual FNDs**

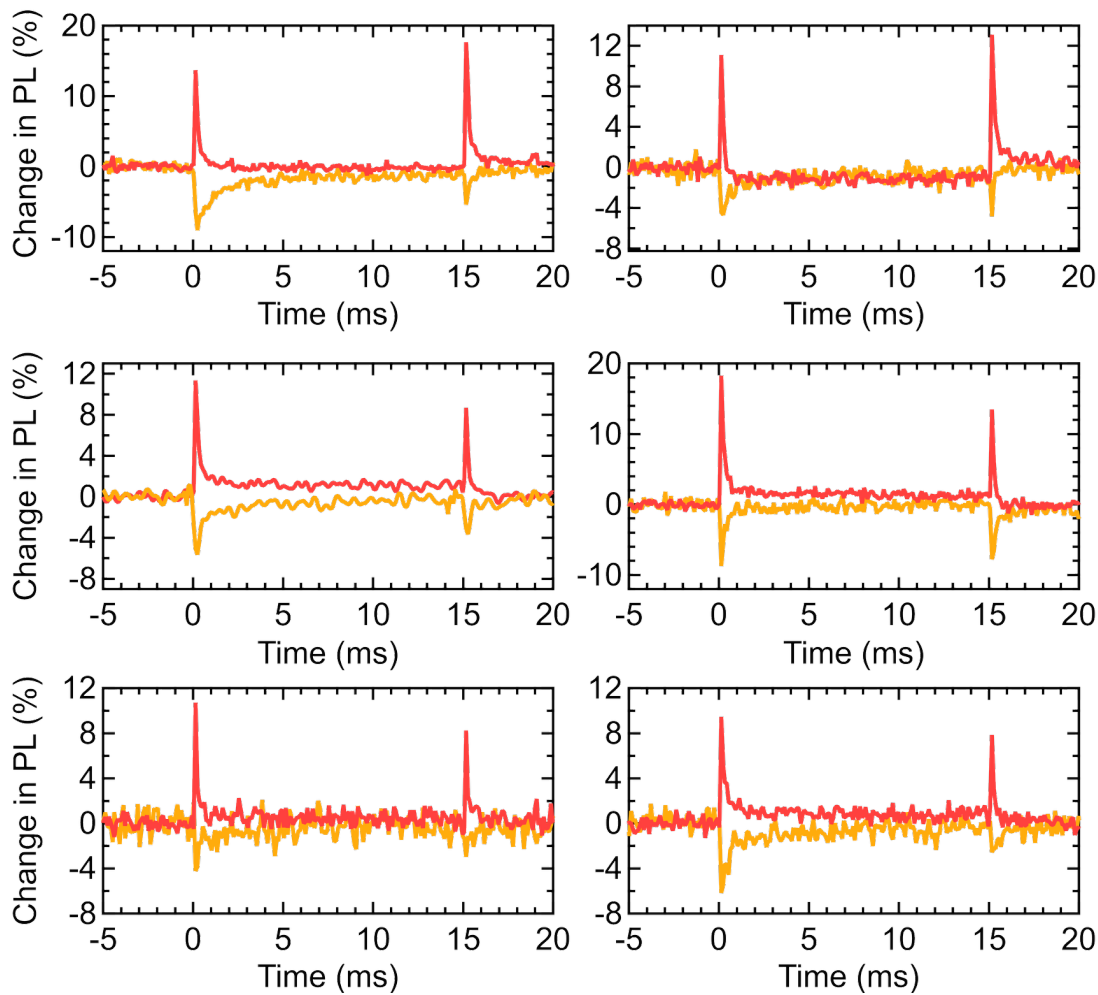

**Figure S7:** Change in NV⁻ (red trace) and NV⁰ (orange trace) PL intensity as a function of time for six different nanodiamonds as an external electric field (100 V) is switched on (t=0 ms) and off (t=15 ms). NV⁻ and NV⁰ PL were acquired in the spectral bands 700-850 nm and 550-650 nm, respectively, using 532nm excitation (~150 µW). These traces illustrate similarities and differences in individual FND's responses to an applied voltage. All traces show a positive peak for NV⁻ PL and a negative peak for NV⁰ PL when the voltage is switched on and a second one when the voltage is switched off. The change in NV⁻ PL is always greater than that in NV⁰ PL. Absolute and relative peak amplitudes vary greatly between individual particles and so does the steady-state PL signal in the 5-15 ms region.

## Voltage-dependence of PL decay

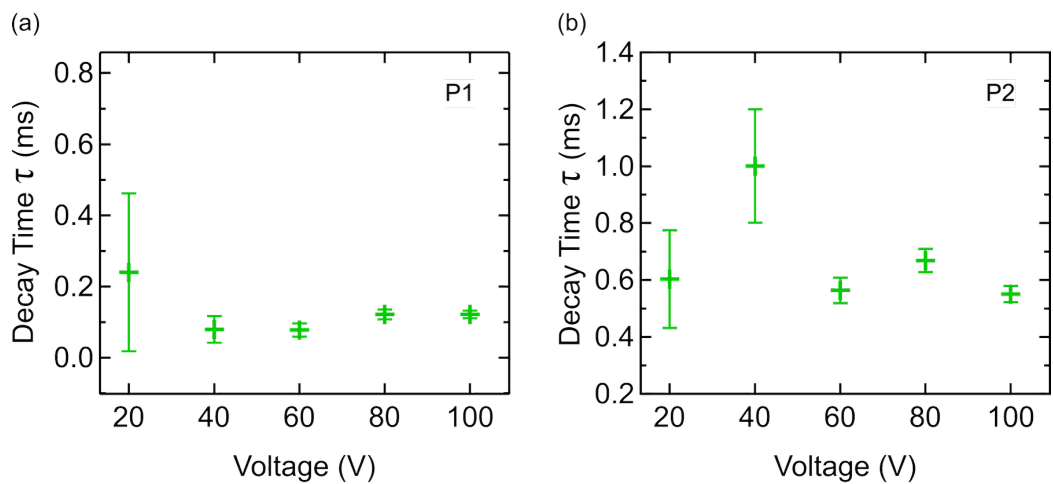

**Figure S8:** Decay time (τ) as a function of applied voltage, obtained from the same time traces analyzed in Figure S6 and Figure 3a of the main text. τ values are extracted from stretched exponential fits to the ΔPL decay curves (Figure 3a) following the initial application of the electric field. Error bars represent the standard deviation in τ derived from the fitting procedure. The large uncertainties at lower voltages are attributed to the low signal-to-noise ratio at low voltages. Both particles exhibit no apparent voltage dependence in the decay rate of ΔPL.

## Voltage-dependence of ΔPL$_{SS}$

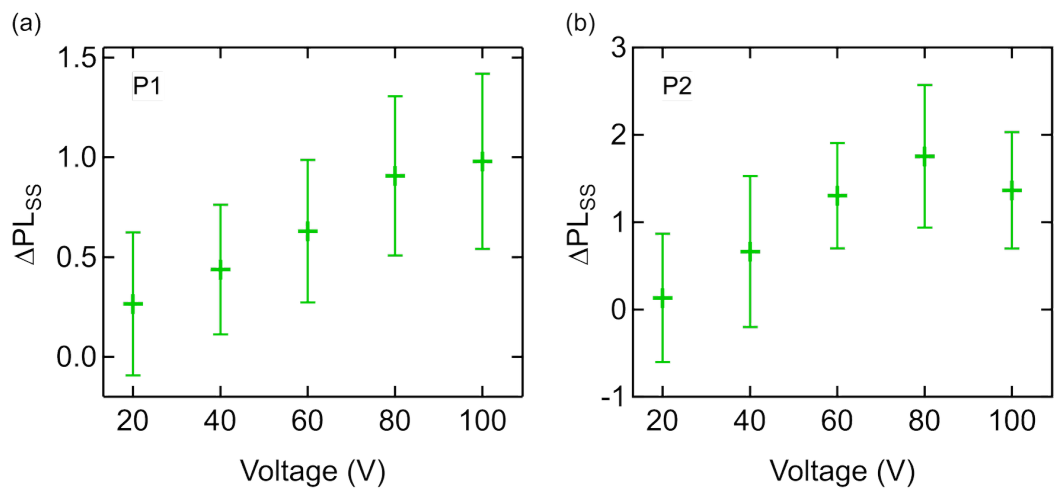

**Figure S9:** Change in PL as a function of applied voltage after 4 ms of continuous electric field application. The plots show the average ΔPL between 4 ms and 14 ms from the relevant time traces, corresponding to the period following the initial PL decay after application of the electric field. Error bars represent the standard deviation of ΔPL within this time window. This average value is referred to as the steady-state PL (ΔPL$_{ss}$), as no significant decay is observed during the prolonged electric field application (see Figure S4). The data are obtained from the same time traces analyzed in Figure S6 and Figure 3a of the main text and reveal a strong correlation between the applied voltage and ΔPL$_{ss}$.

# The effect of the NV$^0$ and NV$^-$ spectral overlap on the observed PL intensity changes

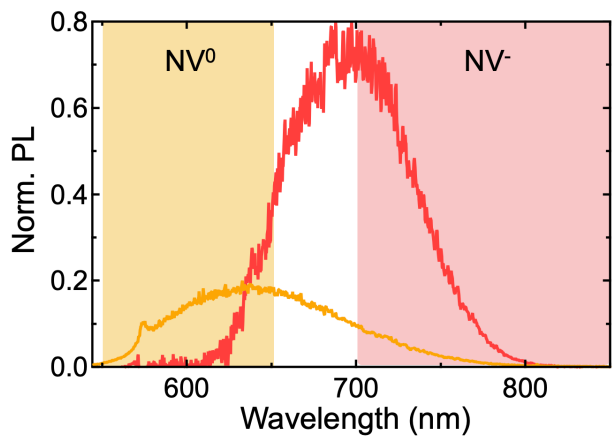

**Figure S10:** Normalized 'pure' NV$^0$ and NV$^-$ PL spectra with a relative intensity of 20% (NV$^0$) and 80% (NV$^-$). Orange and red shaded areas indicate the spectral bands, where NV$^0$ and NV$^-$ PL were collected. A simultaneous 10% increase in NV$^-$ PL and 10% decrease in NV$^0$ PL, will lead to a 8.0% signal increase in the NV$^-$ spectral band and a 3.9% decrease in the NV$^0$ spectral band. Hence, assuming a direct conversion from NV$^-$ to NV$^0$ and assuming identical PL quantum yield and photon absorption cross section, we would expect the absolute value of the amplitude of the NV$^0$ peak in our experiments to be 49% (3.9/8.0) of that observed in the NV$^-$ spectral band. However, this value is sensitive to the initial charge state ratio. For example, assuming an initial charge state ratio of 30% (NV$^0$) and 70% (NV$^-$), the ratio will increase to 87%.

# 1D charge distribution model of a hydrogenated nanodiamond

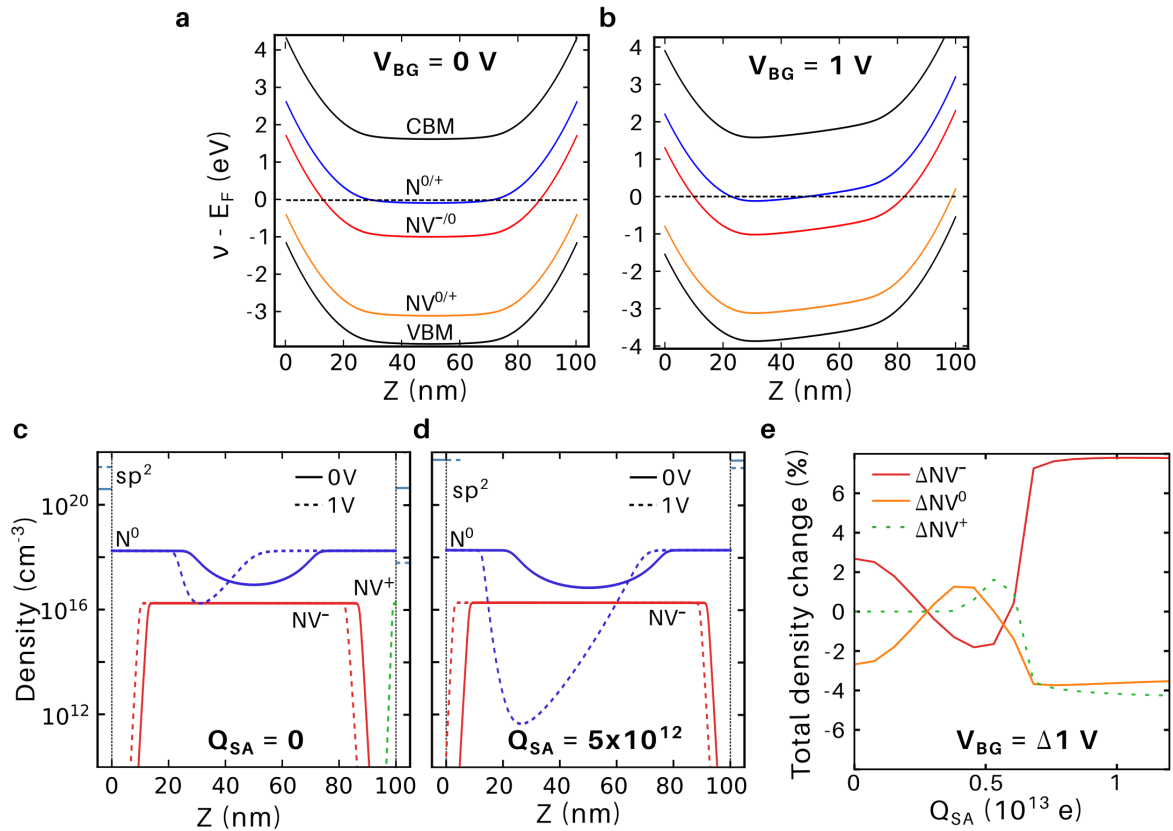

**Figure S11:** a) 1D solved bands for a 100 nm thick section of diamond, there are no surface acceptors ($Q_{sa}=0$), a potential difference is set between the two surfaces of $V_{bg} = 0V$ and b) $V_{bg} = 1V$. c) Key defect charge state densities in the diamond bulk as a function of depth for a surface with $Q_{sa} =0$ d) and $Q_{sa} = 5e12$. e) The relationship between $Q_{sa}$ and the change in total integrated NV charge state density due to a potential difference of $V_{bg} = 1$ V being applied.

## Device sensitivity

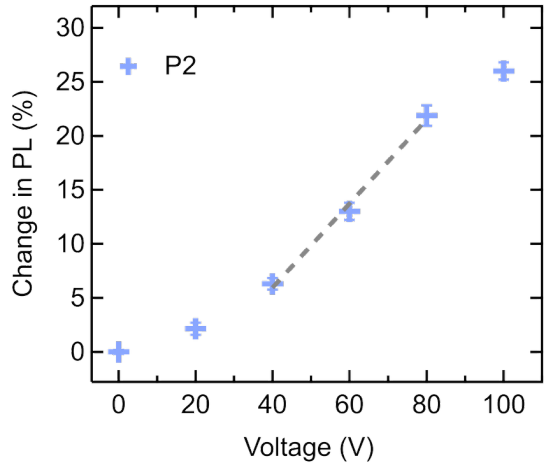

**Figure S12**: Change in NV⁻ PL as a function of applied voltage for FND1 shown in Figure 3 b) in the main text. A linear fit (dashed line) yields a slope of $g = 0.39$ V$^{-1}$, corresponding to $g = 0.06$ mV$^{-1}$cm in our devices. In the shot-noise limit, the signal-to-noise ratio (SNR) is given by

$$SNR = \frac{g\, I_{PL} \Delta E}{\sqrt{I_{PL}}}$$

where $I_{PL} = 750\,000$ s$^{-1}$ is the number of photocounts above 700 nm for a bright FND in our devices and $\Delta E$ is the change in electric field. Assuming SNR=1 for the smallest detectable electric field change leads to

$$\Delta E = \frac{1}{g\sqrt{I_{PL}}}$$

for $g = 0.06$ mV$^{-1}$ cm and $I_{PL} = 750\,000$ s$^{-1}$ this yields $\Delta E = 19$ V cm$^{-1}$ Hz$^{-½}$.

The same calculation for a single NV center with 200 000 cps of PL collected across the entire spectrum in an optimized system [4] and assuming the same slope and spectral filtering ($I_{PL} = 50\,000$ s$^{-1}$ above 700 nm) as above, yields $\Delta E = 72$ V cm$^{-1}$ Hz$^{-½}$.